\documentclass[3p,times]{elsarticle}
\usepackage{amssymb}
\usepackage{lipsum}

\journal{Nuclear Physics A}

\begin{document}

\begin{frontmatter}

\title{Electromagnetic Probes of the Quantum Chromodynamical Plasma}

\author[first]{Gojko Vujanovic}
\affiliation[first]{organization={Department of Physics, University of Regina},
            addressline={3737 Wascana Parkway}, 
            city={Regina},
            state={Saskatchewan},
            postcode={S4S 0A2}, 
            country={Canada}}

\begin{abstract}
In relativistic heavy-ion collisions, electromagnetic (EM) radiation has been used as a sensitive probe of Quark Gluon Plasma (QGP) properties, owing to the smaller EM coupling relative to QCD coupling. To better understand the constraining power of EM emissions on transport properties of the QGP, a deeper understanding of both the theory and phenomenology of EM signals is required. A selection of recent developments in those two areas of QGP EM probes is discussed, with an outlook on how Bayesian model-to-data comparisons can help further quantify our understanding of QGP transport coefficients.
\end{abstract}


\end{frontmatter}

\section{Introduction}\label{sec:intro}
A crucial goal of relativistic heavy-ion collision experiments is to expose the underlying degrees of freedom in the nucleus, the quarks and gluons, into a deconfined state of matter called the Quark Gluon Plasma (QGP). While gluons are solely charged under QCD, quarks carry all three charges of the standard model, including electromagnetism. As the electromagnetic (EM) coupling is much smaller than the strong coupling, once produced, radiated photons can leave the QGP medium with negligible subsequent interaction. Thus, precise information about the state the QGP is encoded within the emitted (real or virtual) photons, which are subsequently detected.\footnote{The high-energy phenomena known as jets are also sensitive to the underlying state of the nuclear medium as they traverse the QGP. However, the usefulness of jets is somewhat limited, owing to the collision energy required for their production.} As real and virtual photons are sensitive to the entire history of the QGP, regardless on the energy between colliding nuclei, their precise detection allows us to learn about transport coefficients of the QGP, such as its viscosity. While real photons are detected directly, virtual photons are picked up through their main decay channel into lepton pairs (or dileptons). 

There are two categories of photon sources during the evolution of the nuclear medium. At high temperatures ($T\gtrsim 200$ MeV), photons from parton interaction are the main source \cite{Kapusta:2023eix}. As the nuclear medium cools, quarks and gluons recombine into hadrons, and the main source of photons stems from in-medium hadron-hadron scatterings and hadronic decays \cite{Rapp:1999ej,Turbide:2003si}. The former continues as long as the medium is dense enough (i.e. the mean free path between collisions is smaller that the overall size of the medium), while the latter persists as long as there are unstable particle decays. 

Among the two sources of electromagnetic interaction, virtual photons, or dileptons, are arguably more interesting as they possess an addition degree of freedom: the center of mass energy of the lepton pair known as the invariant mass $M$ of the virtual photon. The main contribution to dilepton emissions at $T\lesssim 200$ MeV stems from direct decays of low mass vector mesons, i.e. $\rho,\omega$ and $\phi$, that dominate virtual photon production spectrum in the low invariant mass sector $M\lesssim 1$ GeV \cite{Vujanovic:2019yih}. The nuclear medium will induce modification to invariant mass distribution of vector mesons -- i.e. to the vector meson spectral function -- and these modifications are transferred onto the electromagnetic spectral function of the nuclear medium, which peaks around the measured masses of vector mesons. The in-medium EM spectral function contains information about how chiral symmetry is restored as $T\sim 200$ MeV is crossed during the evolution of the nuclear medium created in heavy-ion collisions \cite{Kapusta:2023eix}. As chiral symmetry is restored,it is anticipated that any prominent features of the EM spectral function, such as peaks around the masses of vector mesons, are washed away at high temperatures \cite{Rapp:2009yu}. The featureless dilepton spectrum is present in the intermediate invariant mass range $1\lesssim M \lesssim 2.5$ GeV \cite{Laine:2013vma,Ghisoiu:2014mha}, while additional peaks in the dilepton spectrum appear at high invariant masses $M\gtrsim 2.5$ GeV, which stem from decays of heavy quarkonia into dileptons. Depending on the temperatures reached by the QGP in heavy-ion collisions, it is theoretically possible for quarkonia to melt and be regenerated, and Ref.~\cite{Andronic:2024oxz} explores these possibilities. 

Unlike dileptons, photons do not have any features in their momentum spectrum, both in the hadronic and partonic sectors. Photon radiation from the partonic sector dominates at intermediate photon transverse momenta $1\lesssim p^\gamma_T \lesssim 4$ GeV, while hadronic processes outshine partonic reactions when $p^\gamma_T \lesssim 1$ GeV \cite{Paquet:2023vkq}.     

\section{Production rates and electromagnetic spectral functions}
The electromagnetic spectral function $\Pi_{EM}$ is related to the photon/dilepton (four-)momentum $q^\mu$ dependent rate $R$ via \cite{Geurts:2022xmk}:
\begin{eqnarray}
    E_q\frac{d^3 R_\gamma}{d^3 q} &=& -\frac{\alpha_{EM}}{\pi^2}\frac{\frac{1}{2}g_{\mu\nu}{\rm Im}\left[\Pi^{\mu\nu}_{EM}(M=0,{\bf q};T)\right]}{e^{\beta q\cdot u}-1}\nonumber\\
    \frac{d^4 R_{\ell^+\ell^-}}{d^4 q} &=& -\frac{\alpha_{EM}}{\pi^3 M^2}\frac{\frac{1}{3}g_{\mu\nu}{\rm Im}\left[\Pi^{\mu\nu}_{EM}(M,{\bf q};T)\right]}{e^{\beta q\cdot u}-1}
\end{eqnarray}
where $\alpha_{EM}\approx1/137$ is the EM fine-structure constant, $M^2=E^2_q-|{\bf q}|^2$, $\beta=1/T$, $T$ is the temperature and $u^\mu$ is the flow velocity of the nuclear medium. 

In thermal equilibrium, much understanding has been gained about the EM spectral function, using both perturbative and non-perturbative means. Beyond this, non-equilibrium corrections to the EM spectral functions have also been considered, especially those related to monopole and quardupole off-equilibrium deformations associated with bulk and shear viscous effects, respectively, present in the hardronic and partonic EM production sectors. These two topics are now discussed in turn. 

\subsection{Electromagnetic spectral function in thermal equilibrium}
In addition to non-perturbative calculations based on lattice Quantum Chromodynamics (QCD), the electromagnetic spectral function has recently been extended from leading order to next-to-leading (NLO) order using perturbation theory. This was first done for direct photon production (i.e. at $M=0$) in Ref.~\cite{Ghiglieri:2013gia} before being extended to virtual photons (i.e. $M\neq 0$) \cite{Laine:2013vma,Ghisoiu:2014mha,Ghiglieri:2014kma,Jackson:2019mop,Jackson:2019yao,Ali:2024xae}. While the overall agreement between lattice calculations and NLO perturbative QCD (pQCD) calculations is remarkable \cite{Ali:2024xae}, allowing  to fully appreciate non-perturbative effects, it is interesting to note that a comparison between pQCD and lattice calculations is better for quenched lattice calculations compared to full QCD lattice calculations. This stems in part from the higher temperature present in the quenched calculations, as compared to full lattice QCD. Furthermore, the NLO pQCD calculations have been extended to media with non-zero net baryon number (or baryon chemical potential), which together with hydrodynamical calculations containing a non-vanishing net baryon current, allowed to establish a tight correlation between the real temperature of the QGP from hydrodynamical calculations and the extracted temperature from the slope of measured intermediate invariant mass dilepton yield \cite{Churchill:2023zkk,Churchill:2023vpt,Churchill:2023hog}.

Using hardonic degrees of freedom, two approaches have been undertaken: while one \cite{Hohler:2013eba} relies on chiral effective Lagrangians, the other \cite{Geurts:2022xmk} employs Funcional Renormalization Group (FRG). Using chiral effective Lagrangians, first signs of chiral symmetry restoration have been found by examining the spectral functions of both the $\rho$ mesons and its chiral partner the $a_1$. At $T=170$ MeV, the spectral functions of the chiral partners overlap \cite{Hohler:2013eba}, suggesting chiral symmetry restoration, though more work needs to be done to show how reliable this result is by having other results from different perturbation theory based approaches. Using non-perturbative FRG with mesonic degrees of freedom, similar signs of chiral symmetry restorations have emerged at a higher temperature: $T=300$ MeV \cite{Jung:2016yxl}. Once baryonic excitations are included in FRG, perhaphs this temperature can be lowered to be closer to the pseudo-critial $T_{pc}=156\pm 1.5$ MeV reported by lattice QCD \cite{HotQCD:2018pds}.

\subsection{Extending electromagnetic emission rates to include off-equilibrium contributions}
Hydrodynamical simulations of the QGP have been shown to reliably describe a plethora of experimental observables, provided that shear and bulk viscous pressures are included in addition to the in-equilibrium, i.e. thermodynamics, pressure. While in thermal equilibrium the pressure is an isotropic quantity, dissipative processes giving rise to bulk and shear viscous pressures must deform thermodynamics distribution functions present EM radiation rate. 

In the case of $2\to 2$ scattering producing a real photon with momentum $q$ and using the leading order matrix element $M_{2\to2}$, the in-equilibrium rate is given by \cite{Kapusta:2023eix}
\begin{eqnarray}
\frac{d^3 R_\gamma}{d^3 q} &=& \int \frac{d^3 p_1}{(2\pi)^3 2E_1} f^{(0)}_1 \frac{d^3 p_2}{(2\pi)^3 2E_2} f^{(0)}_2 \frac{d^3 p_3}{(2\pi)^3 2E_3} \times \nonumber\\ 
                           &\times& \left[1\pm f^{(0)}_3\right] \frac{|M_{2\to 2}|^2}{(2\pi)^3 2E_q} (2\pi)^4 \delta^{(4)}\left(p_1+p_2-p_3-q\right)\nonumber\\
\end{eqnarray}
where $f^{(0)}_1=f^{(0)}(\beta[p_1\cdot u])$ and $f^{(0)}_2=f^{(0)}(\beta[p_2\cdot u])$ are incoming particle distribution functions, which are in thermal equilibrium and thus follow the Bose-Einstein/Fermi-Dirac distributions\footnote{Note that $\beta=1/T$ and $u^\mu$ are the inverse temperature and flow velocity of the QGP, respectively.} while $f^{(0)}_3=f^{(0)}(\beta[p_3\cdot u])$ is the outgoing particle distribution that is not the photon.\footnote{Note that since the photon is not in thermal equilibrium with the QGP, Bose-enhancement or Pauli-blocking present in $1\pm f^{(0)}(\beta[q\cdot u])$ doesn't play a role, and thus the photon distribution function is simply 1.} 

Performing a multipole expansion around thermal equilibrium by replacing $f^{(0)}_i\to f^{(0)}_i+\delta f_i$, and assuming that the correction $\delta f$ is smaller than the thermal contribution $f^{(0)}$, Refs~\cite{Dion:2011pp,Paquet:2015lta,Vujanovic:2013jpa,Vujanovic:2019yih} have obtained the monopole and quadripole deformations accounting for bulk and shear viscous pressures, respectively, for photon and dilepton production within two expansion schemes: the  Grad 14-moment \cite{Grad:1949zza} and Chapman-Enskog \cite{chapman1990mathematical} approximations. 
\section{Dilepton and photon production in nuclear collisions}
While photon and dilepton spectra have been shown to be a good thermometer \cite{Churchill:2023zkk,Rapp:2014hha,Shen:2013vja,STAR:2024bpc} and viscometer \cite{Vujanovic:2019yih,Paquet:2015lta,Vujanovic:2017psb,Shen:2013cca}, recent advances have investigated the sensitivity of EM probes to other properties of the nuclear medium such as shear relaxation time \cite{Vujanovic:2016anq}, or the nuclear equation of state. Dilepton have shown sensitivity to the presence of a first order phase transitions in the QCD equation of state, which slows down the expansion rate around the temperature close to the first order phase transitions, yielding a significant increase in the dilepton yield (of about a factor of two) compared the system without a phase transition \cite{Seck:2020qbx}. 

Furthermore, photon and dilepton production has been used to test how quarks are dynamically created following the nucleus-nucleus collision, starting from a system that is predominantly gluonic at early times. Indeed, the nuclear parton distribution functions, giving the probability of finding a quark or a gluon in the nucleus, have shown that at large collisions energies, gluons are the dominant species of particles present in the nucleus and thus the medium created right after a heavy-ion collision is mostly gluonic. However, as the system evolves and thermalizes and entropy increases, quarks start to contribute significantly to the particle population comprising the nuclear medium. In fact, once thermal equilibrium is reached, the rate of quark being converted into a gluon is smaller that the reverse. The dynamical production of quarks in the expansion of the QCD plamsa affects photons \cite{Gale:2021emg} and dileptons \cite{Wu:2024pba}, whose yield and anisotropic flow can be used to pin down the generation rate of electric charge in the QGP.       

Jet-related transport coefficients are another way of understanding the QGP. Recently, the transverse momentum diffusion coefficient $\hat{q}$ of jet partons in nuclear media has been given particular attention, being quantitatively constrained using Bayesian analysis of hadronic observables \cite{JETSCAPE:2024cqe}. Jet-medium interactions can also give rise to photon and dilepton production, allowing to use these EM probes to study how jets exchange energy-momentum with the QGP. While real and vitrual photons stemming from jet-medium interactions are directly sensitive to $\hat{q}$, measurements of lepton pairs have another experimental source that is present, namely semi-leptonic decays of open heavy flavor \cite{Vujanovic:2013jpa,Song:2018xca}. The latter can be removed using heavy flavor trackers \cite{Vujanovic:2017psb} provided sufficient statistics, which is a challenge for many relativistic heavy-ion experiments, thus weak decays of open heavy flavor and antiflavor often contribute to the experimentally reported dilepton signal. The path taken by the heavy quark and antiquark through the QGP will be different and dileptons will be sensitive to this, unlike other open heavy flavor measurements that don't keep track of the open heavy flavor hadron pair. Thus dileptons provide an additional sensitivity to $\hat{q}$ in the intermediate invariant mass range $1\lesssim M \lesssim 2.5$ GeV where diffusion of heavy quarks in the QGP give rise to a non-vanishing dilepton anisotropic flow, such as $v_2$, in that invariant mass range. A first calculation of dilepton $v_2$ using dissipative relativistic hydrodynamics has been reported in Ref.~\cite{Vujanovic:2013jpa}, and a measurement of dilepton $v_2$ by the ALICE collaboration at the LHC is highly anticipated \cite{ALICE:2022wwr}. Indeed, such a measurement can be used to constrain both $\hat{q}$ as well as QGP viscosities. The JETSCAPE Framework allows to revisit that study using a multi-stage evolution of heavy quarks in the QGP \cite{JETSCAPE:2022hcb}.  

As far as jet-medium photon production is concerned, a recent calculation \cite{Yazdi:2022cuk} has shown that about 30\% of photons in the $5\lesssim p_T\lesssim 8$ GeV. Measuring $\hat{q}$ using jet-medium photons allows to put direct constraints on $\hat{q}$ by avoiding hadronization effects. 

As the energy (i.e. $\sqrt{s_{NN}}$) of heavy ion collisions is lowered, the only EM probes can sample, tomographically, the evolution of the nuclear medium. Beyond possible first order transition effects, hydrodynamical simulations of such collisions become less reliable, so a Boltzmann transport approach is used instead. Photon \cite{Gotz:2021dco} and dilepton \cite{Hirayama:2024knv,Staudenmaier:2017vtq,Endres:2016tkg} production can still be computed via Eq.~(2), however the off-equilibrium medium evolution will provide more general $f_i$ distributions than hydrodynamical $\delta f$ viscous corrections, and thus affecting photon and dilepton production. A recent calculation \cite{Gotz:2021dco} of photon production in high-energy heavy-ion collisions uses a combination of hydrodynamical evolutions followed by Boltzmann transport through SMASH. Within that setting, the overall off-equilibrium effects on photon production can be well captures by dissipative hydrodynamics; a result that confirms the use of the hydrodynamical and Boltzmann transport paradigm for high-energy heavy-ion collisions. 

At lower collision energies where the system is anticipated to be further away from equilibrium (with only the highest density region creating the QGP) photons and dileptons from Boltzmann transport are expected to play a more prominent role. A series of lower $\sqrt{s_{NN}}$ experiments will be coming online within the next decade, such as the FAIR CBM experiment \cite{Agarwal:2023otg} and the NA60+ experiment \cite{NA60:2022sze}, and thus the Boltzmann transport approach to EM production will play a more prominent role to explain the EM spectra from far off-equilibrium nuclear media, with hydrodynamical production of EM radiation playing a more secondary role. The inclusion of dileptons together with photons in Boltzmann transport approaches, as has been done in \cite{Gotz:2021dco,Hirayama:2024knv}, will allow to better compare simulations with data, once the latter is available.

\section{Summary and conclusions}
Electromagnetic probes, though rarely produced, have the distinct advantage of avoiding any hadronization effects, thus making them an invaluable probe for directly extracting nuclear medium properties. In the case of high-energy nucleus-nucleus collisions, properties like the nuclear equation of state, its viscous transport coefficients, as well as jet-related transport coefficients such as $\hat{q}$ are more readily available to be constrained. To quantitatively extract nuclear matter properties, a Bayesian analysis combining EM radiation with other nuclear media observables is anticipated. This multi-messenger approach together with better understanding of theoretical systematic uncertainties, such as those associated with $\delta f$ calculations, is to be included within the next generation of Bayesian model-to-data comparisons to generate a more robust understanding of the nuclear medium. 

Furthermore, precise Bayesian model-to-data comparisons allow to put tight constraints on fundamental phenomena of nuclear medium transition that is not accessible through direct observation. One such property chiral symmetry restoration, that ideally would require direct observation of how the spectral function of chiral partners --- e.g., $\rho$ and $a_1$ mesons --- behaves as temperature is increased. While the in-medium $\rho$ spectral function is accessible through dileptons, given that the vector nature of the $\rho$ meson allows a coupling to EM radiation, the spectral function of the $a_1$ chiral partner isn't easily accessible, as the axial-vector nature of that meson prevents its coupling to real or virtual photons \cite{Rapp:2009yu}. However, more precise measurements of dileptons together with Bayesian analysis that include theoretical systematic uncertainties, can be used to ascertain whether experimental data favors calculations containing chiral symmetry restorations effects. Indeed, Bayesian model selection provide us with a quantitative measure to discriminate one model calculation versus another.

Finally, Bayesian analyses can advise areas where better measurements of a particular observable can provide substantial constraint on a given nuclear matter property \cite{JETSCAPE:2020mzn,Heffernan:2023utr}, thus increasing the synergy between theory and experiment. A Bayesian meta-analysis --- including EM probes, soft hadronic observables, and jet-related measurement --- together with a detailed account of theoretical and experimental systematic uncertainties are need in a holistic approach to understanding QGP properties.

\section*{Acknowledgements}
The work was supported in part by the Natural Sciences and Engineering Research Council of Canada, reference number SAPIN-2023-00029, by the University of Regina President's Tri-Agency Grant Support Program, and by the Canada Research Chair program, reference number CRC-2022-00146.

\bibliographystyle{elsarticle-num} 
\bibliography{example}

\end{document}